# Half a cubic hectometer mooring array of 3000 temperature sensors in the deep sea

by Hans van Haren[*], Roel Bakker, Yvo Witte, Martin Laan, Johan van Heerwaarden

NIOZ Royal Netherlands Institute for Sea Research, P.O. Box 59, 1790 AB Den Burg, the Netherlands.
*e-mail: hans.van.haren@nioz.nl

ABSTRACT

The redistribution of matter in the deep-sea depends on water-flow currents and turbulent exchange, for which breaking internal waves are an important source. As internal waves and turbulence are essentially three-dimensional '3D', their dynamical development should ideally be studied in a volume of seawater. However, this is seldom done in the ocean where 1D-observations along a single vertical line are already difficult. We present the design, construction and successful deployment of a half-cubic-hectometer (480,000 m$^3$) 3D-T mooring array holding 2925 high-resolution temperature sensors to study weakly density-stratified waters of the 2500-m deep Western Mediterranean. The stand-alone array samples temperature at a rate of 0.5 Hz, with precision <0.5 mK, noise level <0.1 mK and expected endurance of 3 years. The independent sensors are synchronized inductively every 4 h to a single standard clock. The array consists of 45 vertical lines 125 m long, at 9.5 m horizontally from their nearest neighbor. Each line is held under tension of 1.3 kN by a buoyancy element that is released chemically one week after deployment. All fold-up lines are attached to a grid of cables that is tensioned in a 70 m diameter ring of steel tubes. The array is build-up in harbor-waters, with air filling the steel tubes for floatation. The flat-form array is towed to the mooring site under favorable sea-state conditions. By opening valves in the steel tubes, the array is sunk and its free-fall is controlled by a custom-made drag-parachute reducing the average sinking speed to 1.3 m s$^{-1}$ and providing smooth horizontal landing on the flat seafloor.

**1. Introduction**

Transport of matter in geophysical environments occurs on scales over a range of some nine orders of magnitude. At present, this range of scales poses an unsurmountable challenge for scientists studying transport and exchange processes, whether they are theoreticians, modelers or observers. Yet, knowledge about the redistribution of heat, matter and nutrients is important for life. Even studies with limited resolution of scales over two to three orders of magnitude challenge scientists, because these require such resolution in 2 or 3 spatial dimensions '2D' or '3D' besides time, so that choices for the range of investigation have to be made. While such choices are nowadays perhaps easier made for sophisticated numerical models, they are difficult for observational studies. For example, the important turbulent diapycnal exchange in density stratified ocean and atmosphere is mainly sampled quantitatively in 1D, via a moored string or lowering/lifting of instruments, qualitatively in 2D, via remote sensing techniques, while turbulence processes are essentially 3D (e.g., Tennekes and Lumley 1972). Thus, as stated in a review paper (Sun et al. 2015), major progress in ocean and atmospheric (boundary layer) exchange studies is expected from observations in 3D.

Focusing on the ocean, the daytime solar heating stores large amounts of potential energy near its surface that become distributed into the interior via kinetic processes (Munk and Wunsch 1998). The heating from above imposes a 2D character on the ocean interior, which becomes stably density stratified in the vertical because warm water is less dense than cold water. As the ocean, like other geophysical environments, has very large length scales resulting in large Reynolds numbers, motions are nearly everywhere turbulent. Thus, turbulence dominates vertical exchange processes, even when the environment is stably stratified in density. With the stable density stratification, large vertical gradients of other constituents like oxygen, nutrients and suspended matter can be expected. Stratification reduces the vertical length scale of turbulence, resulting in quasi-2D anisotropic 'stratified turbulence', but does not block turbulent exchange. As a result, sites of enhanced diapycnal exchange are worth studying for knowledge on the exchange of constituents.

In the ocean interior outside the wind-driven surface wave influence, most turbulence is generated via the breaking of internal waves, which are supported by the stratification, above sloping underwater topography (e.g., Eriksen 1982; Thorpe 1987). Due to severe limitations of the harsh ocean conditions imposed by salinity, pressure and waves, 3D deep-ocean observations on internal wave turbulence dynamics are few and far between and much is still to be learned about the evolution characteristics of relevant processes that dominate diapycnal turbulent exchange via internal wave breaking.

Ocean internal waves can attain vertical amplitudes exceeding 100 m, which develop the largest energy containing turbulence scale $L_e \sim O(100)$ m (e.g., Farmer et al. 2011). The Ozmidov scale $L_O \sim O(1)$ m is the largest scale of isotropic turbulence in a stratified environment (Ozmidov 1965), while the smallest turbulence energy containing scale dissipating into heat is the Kolmogorov scale $L_K \sim O(0.001)$ m. As the five orders of magnitude between $L_e$ and $L_K$ are not resolvable by present-day ocean turbulence observational techniques, focus has been on either resolving $L_K$ to $L_O$, mainly using microstructure profilers (e.g., Oakey 1982; Gregg 1989), or resolving $L_O$ to $L_e$, using Conductivity-Temperature-Depth profiling (Thorpe 1977; Dillon 1982). Both techniques use instrumentation that is lowered from a ship in a quasi-stationary, semi-Eulerian manner, with a limited coverage in time.

Higher resolution in time can be obtained either by following water particles in a Lagrangean manner, or by mooring instrumentation fixed in underwater space and observe the flow going past in a Eulerian manner. As Lagrangean observations are still not feasible for studying ocean dynamics, moorable Eulerian systems are used to avoid Doppler shift in frequency in studying ocean internal wave turbulence (Gerkema et al. 2013). Thus, $L_O$ to $L_e$ have been observed using 1D moored temperature sensor strings, of 10 to 20 m vertical separation at 0.01 Hz sampling rate (Aucan et al. 2006; Levine and Boyd 2006), and of 1 m vertical interval separation at 1 Hz sampling rate using high-resolution temperature 'T-'sensors (van Haren and Gostiaux 2012).

Following limited 3D-moored observations using conventional (non-turbulence measuring) instrumentation (Briscoe 1975; Pinkel 1975; Saunders 1983), a prototype small-scale 3D mooring array for studying internal wave turbulence was constructed consisting of five parallel lines equipped with a total of 500 high-resolution temperature sensors (van Haren et al. 2016b). The mooring lines were 100 m tall but only 4 m apart, so that only $L_O$ was partially resolved in 3D. Nevertheless, statistical information was obtained on the transition from anisotropic 2D stratified turbulence to isotropic 3D turbulence through a non-smooth inertial subrange (van Haren et al. 2016a). Technically, this prototype of multiple parallel lines mooring array could only be deployed in a single overboard deployment operation, for which the lines were fold-up. However, its size did not fill a 3D volume $O(L_e)$ as the horizontal spatial scale resolution was limited, and thus far no large-scale 3D internal wave – turbulence observational study has been made in the deep sea.

In this paper, we present the design, construction and deployment of a large 3D mooring array for studying development characteristics of deep-sea internal wave turbulence at $L_O$ to $L_e$ scales in 3D as in a large fluids laboratory. To fill a volume of (half) a cubic hectometer (1 cubic hectometer = 1 Mm$^3$) of deep seawater with about 3000 high-resolution temperature sensors, we designed a multiple (45) parallel vertical lines array to resolve 2-m vertical scales over a range of 125-m and <10-m horizontal scales. As such a multiple-line-mooring is not standard deployable from a ship, we compacted the entire array in a flat floatable form before sinking it in a semi-controlled free-fall to the seafloor. We let the mooring lines automatically unfold after landing at the seafloor. We describe the technical details, the elaborate construction and the deployment of this large 3D-T mooring array for use in the deep Western Mediterranean where all governing dynamics are O(0.001)°C: A true challenge for our temperature sensors.

**2. Modular design**

Inspired by the instrumental set-up of the cabled network of underwater neutrino telescopes ANTARES (acronym for Astronomy with a Neutrino Telescope and Abyss Environmental Research; Ageron et al. 2011) and especially KM3NeT (acronym for cubic kilometer neutrino telescope; Adrián-Martinez et al. 2016), we sought a means to study ocean dynamics in a volume of deep-sea waters. From our experience with moorable high-resolution NIOZ T-sensors as the simplest means to study ocean dynamics processes (van Haren 2018), it seemed logic to fill a volume of sea water with such sensors. Unfortunately, integration of our T-sensors with the optical detection units of the neutrino telescopes was impossible because obstruction of optical sensors by objects like T-sensors was to be avoided and because the 10 W heat shield of the optical detection units was artificially modifying the local weak temperature variations. Thus a separate 3D mooring array had to be designed for the T-sensors that should extend the prototype-3D five-lines/500 sensor array (van Haren et al. 2016b).

Seeking to launch half a cubic hectometer large mooring array that somehow could be compacted, we borrowed ideas from a fish farm, its floating ring, from beach-sand replenishment, its steel tubes, from a bicycle wheel, its spokes to prevent the ring from excessive flexing, from compacting a mooring line, to become vertical after self-unfolding, from parachutes, to regain control during free-fall sinking to the seafloor. (To prevent too much drag and tension in a mooring line from ship motions due to surface wave action, top-first is the more common way of deploying oceanographic sub-surface moorings.)

The eventual result following scientific, feasibility, budgetary, and other design studies, is a large mooring array '3D-T' consisting of 45 vertical mooring lines that are 125 m tall and 9.5 m horizontally from the nearest neighbor, holding a total of 2925 T-sensors. The mooring lines are attached to a steel cable network that is held inside a 70-m diameter ring of steel tubes that float when filled with air and sink when filled with water. Together with the fabrication of necessary custom-made tools, it was constructed in one year precise and Figure 1 shows the final hour before sinking to its underwater position.

## 3. Technical details, description of components

*a. Temperature sensors and their sticking to a vertical line*

The core of the large 3D-T mooring array consists of modified high resolution standalone NIOZ4 T-sensors. For this project, 3030 new T-sensors have been built, with the same characteristics as previous NIOZ4: 600 bar (6000 m) pressure resistance (depth range), $<0.0001°C$ noise level, $<0.0005°C$ precision, and synchronization via induction to a single standard clock every 4 h so that the individual clocks deviate by <0.02 s (van Haren 2018). The recent modifications involve: Easier to mount T-sensor resistors (NTCs), a new lithium battery slightly larger and with approximately 1.5 longer lifespan than AA-penlight, adapted software so that the synchronization search space is increased by 2 s per week with synchronization in 6 groups of up to 8 mooring lines from a predetermined start date. The group synchronization pulses are separated in time by 0.5 h. The 0.1-s long sampling at a rate of once per 2 s allows an internal data storage and battery life span for at least 4 years.

Over a course of nearly 3 months, all T-sensors were calibrated to a precision of 0.0001°C in a custom-made laboratory bath with a capacity of 200 T-sensors. In order to complete the entire calibration operation in time, the range was limited between 8 and 18°C, which is adequate for deep Mediterranean waters having temperatures around 13°C. After finishing all calibrations, the T-sensors were collectively programmed to start on November 1, 2020 at 06:00:00 UTC. The single synchronizer was restarted with the first synchronization pulse to be sent 4 h before the sensors' start time to avoid mismeasurements due to system startup.

As the synchronization pulse is received by the T-sensors via induction, it is sent through electrically conducting cables that are well insulated from the saline seawater. For this purpose, water resistant XLPE 16 AWG 1.5mm$^2$ conducting copper wires are taped along the steel cable network. The 125 m vertical mooring lines consist of 5 mm diameter steel cable with a breaking strength of 20 kN and a 0.6 mm thick nylon coating for insulation. At the top of every vertical line a brass-alloy functioning as an anode is soldered to the coating-removed steel end. The saline water closes the electric circuit to the synchronizer. The nylon coating of the vertical lines is sturdy and relatively smooth, making the line a good compromise between strength and current drag resistance to minimize mooring motion and to ensure best-possible Eulerian measurements. The coating also holds adhesive tape well, which we use to stick T-sensors to the cable. Per vertical mooring line we stuck 65 T-sensors using yellow Japanese electricians' adhesive tape, 63 sensors to measure temperature at a distance of 2.0 m and 2 sensors to measure temperature/tilt near top and bottom of each line.

*b. The small ring*

As the mooring lines cannot be vertical before the mooring array is at the seafloor because they certainly would entangle each other during the sinking, they need to be compacted. To compactly transport an individual mooring line with T-sensors and have it unfold at the seafloor, a design analogous to the five-line mooring was chosen (van Haren et al. 2016b). Each 125-m long mooring line is looped on a 2.5-m diameter aluminum 'small ring' (Fig. 2). The T-sensors are clicked into pieces of cable channel and are held in place by the overlying top-buoy of which the central shaft is held in two stands at its top and bottom. The 2-m long line-loops between each T-sensor pair are stacked in a roof-tile manner, are pressed into clips and held between tension straps. The filled ring weighs approximately 400 N above water, the mooring line 150 N underwater. The 3000-m rated syntactic foam top-buoy delivers 1450 N of buoyancy and is held by a band with a zinc-aluminum compound chemical release that opens after 5-7 days underwater.

*c. The large ring*

In order to get the 45 small rings at 9.5 m horizontally apart to its nearest neighbor and electrically connected in the deep-sea, they could be lowered on their own anchor under coordination of a high-resolution acoustic positioning system. This would already be a ship-time-consuming work and the line-location would be not more precise than about 1 m horizontally at 2500 m water depth, but, furthermore, the assistance of a deep-sea underwater robot would be needed to electrically connect the network to have all sensors measure simultaneously upon instruction from a single synchronizer. Such sea operations would be difficult, costly and laborious.

It was alternatively decided to electrically connect the mooring lines to the synchronizer at the sea surface and to let the entire compacted mooring array sink in controlled free-fall to the seabed in a single deployment operation. To this end, a large steel-tube ring (henceforth shortened as 'large ring') the size of more than half a football field, almost 70 m in diameter, is strung with a network of 12 mm diameter steel cables (Figure 3). The extra two cables forming the X are crucial for sufficient sturdiness. The network prevents excessive ring-flexing and is a necessary means to attach the vertical mooring lines. Small rings with compacted mooring lines are mounted at each intersection of the steel cable network. The main connection to a vertical mooring line is at the intersection through a hole in the aluminum frame of a small ring. For additional support, clamps fix the two crossing network cables at two opposite sides of a small ring. There are 37 intersections inside the large ring.

As the locations of the remaining 8 small rings would be at the outer-edge of the ring itself, they are shifted and mounted just inside the large ring. This displaces their vertical mooring lines about 1.5 m horizontally from the gridded positions, becoming 8.0 m to their nearest neighbor (Fig. 4). The mounting of the 'corner small rings' is via 3 short sections of cable clamped to two perpendicular main cables. As a result, the corner small rings are less well fixed in the horizontal plain, as they pull the two main cables towards each other by a few centimeters.

The large ring consists of 18 straight steel tubes 12 m long, miter cut at the ends at 10º. Flanges are welded to both open ends to prevent water moving through the entire ring when tilting during filling. The tubes have a diameter of 0.61 m and a wall thickness of 6.3 mm (steel S355). Each tube has manually closeable holes on both sides on both ends. When closed, they float with approximately 22 kN/tube. When full of water they sink at about -12 kN/tube. This means that together they automatically act as anchor weight for all 45 vertical mooring lines. Due to the net 1.3 kN buoyancy of each top-buoy and vertical mooring line, the 12 mm steel cable network is pulled upwards like a reverse trampoline by a total of nearly 60 kN. Cable stretching tests with the same pulling force as through the top-buoys have shown that the highest central point of the network will be approximately 2.5 m above 'ground level'. The ground level is the center of the steel tubes, the level of the attachment points of the 12 mm steel cables at 0.30 m above the seafloor if the bed is solid. Underwater, each cable is directed at a deflection angle of >4.2º to the horizontal at the connection to the steel tubes and provides a pulling force of <16 kN on a flange or a cable mounting saddle around the steel tubes. There are 32 saddles for the 16 steel cables and other fixings. They can be safely loaded up to 27 kN (with a safety factor of 4). The flanges of the tubes are met using 16 M20 bolts per connection. Per lifting point 225 kN can be hoisted.

The M20 bolts per flange-connection are tightened to 450 N/mm$^2$ tension with an impact wrench and checked with a torque wrench. In water, after correction for Archimede force, the large ring weighs 215 kN, a small ring 0.25 kN and the entire 12 mm steel cable network including intersection-connections about 6.5 kN. The total mooring array weighs approximately 177 kN when completely submerged in water and the steel tubes are filled with water. The large ring has a volume of 18x3.4 m$^3$ = 61 m$^3$ and the assembly floats by about 450 kN when the steel tubes are filled with air.

In winter in a 5x5x5 m open-air basin, water filling of the large ring was tested using a 4 m long steel dummy tube with identical diameter and wall thickness as the final 12 m long tubes. At the bottom of the dummy tube 3 holes of 40 mm diameter were made, at the top 3 holes of 12 mm diameter. The diameter of holes at the top determines the speed of air release, *i.e.* the speed of water filling the tube. The diameter of holes at the bottom must be large enough to quickly replace the remaining air with water when the tube is sinking before the ambient pressure implodes the tube. One open top hole fills the tube after about 6 minutes. Two holes fill the 4-m tube in about 3.5 minutes.

For the final design we decided on two 12 mm diameter holes with manual ball valves at the top, so it was expected to take about 10 minutes for filling a 12 m long steel tube. The tube can resist a pressure of 10 bar before implosion. In order to have it full of water within one minute, at an expected free-falling speed of 1.5 m s$^{-1}$ (using Bernouilli law, at an average pressure of 5 bar) before reaching 100 m under the sea surface, the two holes at the bottom need to have a diameter of 60 mm. These holes are closed with removable slid valves.

For verification and for further construction, the large ring was built-up in a field next to NIOZ. All saddles and ample zinc/aluminum alloy anodes were mounted to the steel tubes, and the 12 mm cable network was laid out entirely. The steel cables were verified to be precise in length to within 0.01 m, the large ring was a 70 m diameter circle to within ±0.05 m. Custom-made connections at the cable network intersections were tested and all intersections were precisely measured and marked with paint. The positions of the drag-parachute lines (see Section 3.d) fold-ups were determined and clip-bands fastened to the appropriate steel tubes.

To bring the complete large ring to the mooring location, it is towed at four points of the large ring. The four towing lines consist for the most part of 26 mm diameter Dyneema (660 kN breaking strength), each added with a 15 m long, 44 mm diameter Polyamide stretcher (400 kN breaking strength). The towing from the harbor out to sea is performed by a commercial towing company. At sea the two outer lines are taken in, while the two inner lines are handed to the research vessel via rubber boats.

*d. Drag-parachute*

From the moment the compacted large ring mooring array starts to sink, the stability is out of the system and control is lost. The ring falls on a side, the side of least resistance during sinking through the water column, the ring may flip and may land on a side, break, or tumble upside down. This has been confirmed during a water filling test using a 1:35 scaled aluminum tube model of the large ring and using a 1:35 scaled solid steel model with the same hydrodynamic characteristics during free fall as the large ring. In order to regain control (actually: in order not to lose control permanently) a megaton crane of 70 m height would be necessary, which is impossible. Instead, a drag-parachute is designed and attached to the large ring to stabilize the ring during free fall so that the ring will land smoothly and horizontally on the flat seafloor (Fig. 5). The drag-parachute is attached to 6 points distributed over the large ring ending in a single point below the drag-parachute. The single point attachment is required to avoid destabilization of the drag-parachute that should remain horizontal during the sinking at all times.

The size of the drag-parachute is determined from various tests with the model consisting of 2 m diameter solid steel ring. At a free fall speed of 1.5 m s$^{-1}$, measured and calculated from the surface area and weight of the final large-ring ensemble, the model has a weight of 26 kg and a tube diameter of 26 mm. Two drag-parachutes are tested, one with braking resistance $F_b = 0.1F_g$, i.e. 10% of underwater weight $F_g$, and one with $F_b = 0.2F_g$. The braking effect consists of a buoy, in the model a single air-filled PET bottle of 1.5L, that is combined with a drag resistance disk having diameters of 0.1 and 0.2 m, respectively. In an 8 m deep swimming pool it turned out that both disks work adequately (Fig. 6). The 0.2 m diameter large disk stabilized the ring after a 5-m drop, regardless of the angle of launch of the model ring. The 0.1 m diameter small disk also stabilized, but not completely during the 8-m drop height.

Although the fall in (scaled-up) reality takes at least ten times as long as in the swimming pool so that the small disk would be sufficiently stabilizing the large ring before touching the seafloor, a drag-parachute with braking force of approximately $F_b = 0.15F_g$ was chosen. The actual drag-parachute for stabilizing the large ring during sinking consists of a standard mooring top-buoy and a custom-made drag-parachute. The standard buoy has 3.7 kN lift capacity plus, at a vertical falling speed of 1.4 m s$^{-1}$, 1.7 kN of resistance braking power. This buoy comes 15 m above a newly built large aluminum drag-parachute (Fig. 7). The large drag-parachute has an area of approximately 10 m$^2$, its aluminum frame weighs 2.1 kN underwater, and it carries 48 glass spheres providing a total of 12.2 kN of buoyancy (see Appendix for calculations). This parachute hangs about 55 m above the large ring when it is stable (Fig. 5). The large ring is attached to 70 m long Dyneema lines that are each attached to an IXSea Oceano 2500S acoustic release. To each of the 6 releases an (unsynchronized) temperature/tilt sensor is taped, for monitoring the drag-parachute balancing during the sinking. The 6 lines converge at a single attachment point via a 100 kN swivel below the parachute. The large drag-parachute is separable in three equal parts that fit in a sea-container or on a truck for easy transportation. All 48 glass spheres are protected by plastic caps and interconnected with aluminum strips to avoid dangling during handling, transportation, towing and sinking.

During installation for towing out, the lines of the drag-parachute are lightly secured via tape to the 12 mm network steel cables, and guided over top-buoys of small rings. Via a click system excessive line is folded in loops on top of steel tubes of the large ring. The lay-out of lines is made so that the drag-parachute can be kept afloat on the ship's side, outside of and asymmetrical with respect to the large ring. By semi-controlled opening the water-filling valves of the steel tubes starting on the side of the drag-parachute and, hence, of the ship, the large ring is forced to sink on that side first. No sooner than sinking at least 4 m of the first-opened tubes, the drag-parachute can move inside the ring. Thus, entanglement of the drag-parachute's lightly floating lines with small rings is avoided.

Also for proper drag-parachute-steering of the large ring, Dyneema lines were chosen as they have little stretch (approximately 0.3% for the 16 mm diameter lines, which have a breaking strength of 175 kN). The lines were polyester-coated to prevent too easy cutting when touching sharp objects. After checking, the lines were not manufactured to the same length, and by far not meeting the precision of 0.01 m over 70 m of the 12 mm steel cables in the network: The lines varied in length by up to 0.75 m. Doubled hoisting slings were added so that all 6 'lines' have an equal length with a precision of 0.05 m.

To prevent the lines from getting stuck between the glass spheres of the drag-parachute, a railing was built around the outer ring also to prevent spheres from breaking when the drag-parachute is against the side of the ship or against the large ring in the sea. At full sinking speed, the 6 drag-parachute lines are pulled at 27 kN, which is about half the maximum load of an IXSea standard deep-sea acoustic release. Nevertheless, for safety reasons and for recovery of the rather expensive release-ring, the single acoustic releases per line are 'doubled' with a chain through a large steel ring. Each acoustic release is held in a tandem of 2 glass spheres, so that the ensemble is net buoyant with about 0.25 kN after release.

*e. Chemical release*

After the ring has landed on the seafloor, and the location plus orientation have been accurately determined by triangulating all 6 acoustic releases, the drag-parachute is disconnected and recovered from the sea surface. To avoid entanglement of different lines, this should be done well before the individual temperature sensor mooring lines are released. It is too expensive to equip each of 45 mooring lines with a deep-sea acoustic release. Instead, a chemical release is purchased for each line, an aluminum/zinc alloy that dissolves after 5 to 7 days in seawater at 38 g/kg salinity and a temperature of 13 °C. The in-house tests yielded results in accordance with the specifications of the manufacturer. Each chemical release is linked in a custom-made strap-band that holds the top-buoy of a mooring line. New, the release breaks at a force of exceeding 3.7 kN, which is sufficient to hold the buoy during the towing operation and the sinking of the mooring array.

*f. Assembly raft*

To attach the small rings in the corners of the large ring and at the intersections of the 12 mm steel cables during the build-up in port, an assembly raft has been constructed. This raft weighs about 6 kN and has a buoyancy of about 20 kN. With a small ring and four people, the whole still floats net 10 kN (Fig. 8). Thus on one side, a floating tube can be safely taken out of the water to clear the way after the small ring has been attached to the steel cable network. Pulleys are mounted on Davids to lift the steel cables and to allow the small ring to be lowered. Two Talamex TM48 electric outboard motors provide propulsion of approximately 4 hp (total thrust 2x48 LBS, for a total maximum boat weight of 29 kN).

*g. The mooring site*

The site of the 3D-T mooring array is in the vicinity of the underwater neutrino telescopes KM3NeT-F ('ORCA') and ANTARES (Fig. 9), in French territorial waters of the Western Mediterranean Sea. The telescopes use highly sensitive optical instrumentation attached to multiple mooring lines so that they also sample a volume of seawater, like 3D-T. The technique of deployment of the self-unrolling telescope-lines bears many similarities with that of 3D-T, although the 30 kN mooring lines including anchor weights are lowered via a ship-winch to the seafloor one-by-one. Results from the complimentary optical and temperature sensors yield insight in deep-sea biology activated by variations in their physical environment, including the rapid deposition of fresh material from the surface to the seafloor during late-winter (van Haren et al. 2011).

The site is due south of Porquerolles island, about 40 km southeast of Toulon harbor, in a restricted zone. The seafloor is flat, <0.1° slope to the horizontal, with a substrate of sand and finer deposited materials. The site is less than 10 km south of steep and rugged continental slope topography. As this major topography steers a boundary water-flow, the Northern Current, mainly westward directed but heavily meandering due to its instability, interesting variations in deep-sea (sub-)mesoscale eddies, dense water formation, near-inertial waves, internal wave mixing and vertical transport are expected (Crépon et al. 1982; Albérola et al. 1995; Gorsky et al. 2002). Tidal motions are small in the area. Maximum flow speeds can become 0.35 m s$^{-1}$ at the seafloor, in late winter under enforced Northern Current. The mooring array is constructed so that the associated drag will not displace the top-buoyancy by more than the employed horizontal distance in the steel cable network under the maximum flow speeds. The associated maximum vertical excursion of the top-buoy is thus smaller than 0.15 m. To monitor currents and mooring line deflections, 3 top-buoys are equipped with a single-point AquaDopp acoustic current meter with pressure and tilt sensors. Every line holds two temperature/tilt sensors.

**4. Sea operations**

*a. Preparation*

The entire large 3D-T mooring array is uploaded on four trucks for land transportation. Two trucks each loaded 9 steel tubes and two trucks each loaded two 20-foot sea containers with materials. The mooring array is assembled in floating, flat form in the port of Toulon (France), at the CNIM quay in la Seyne-sur-mer, in 8 days.

A 75 Ton crane is hired to link all 18 steel tubes of the large ring in pairs on the quay. Each pair is put correctly oriented in the harbor waters with the crane at a fixed spot on the quay, and attached at the flanges by the fixing the bolts to the previous pair(s). The assembled partial ring is pushed by rubber boat for the next pair of tubes to be placed.

By means of a hired telehandler the steel cable network is towed into the water via pullies attached to the opposite side of the ring. Securing cables to the large ring is done via the rubber boat. The network-intersections are pre-assembled on the quay and 20L empty water barrels are used as floatation for later fishing the connections to attach the small rings to the steel cable network from the assembly raft.

The telehandler is also used to launch into the harbor waters the rubber boat, the assembly raft and one by one the quay-side-assembled small rings. In the final stage of assembly, the cables of the network are weakly, <0.1 kN, but equally tensioned, the synchronizer electric cabling is fixed to the central small ring, the drag-parachute lines are laid out and, just prior to arrival of the tugboat, the four towing lines are moved from the fixation to quay-side to their towing positions.

*b. Deployment*

On 09 October 2020, a perfect day with very little wind, speeds <3 m s$^{-1}$, and waves, heights <1 m, the takeover of the large ring from the tugboat to the Dutch R/V Pelagia went smoothly. The mooring assembly was held stationary at position east of the target site in a steady westward flow of about 0.5 m s$^{-1}$ at the surface. Two towing lines were disconnected from rubber boats and taken over by the tugboat, after which the R/V Pelagia maneuvered closer and took over the two inner towing lines from the rubber boats. The entire ring was checked prior to deployment. Securing straps of acoustic releases were removed and some repair work was done on the drag-parachute lines. In the large ring, locking pins of all 45 buoys were removed. After the drag-parachute was lowered into the sea from the R/V Pelagia and attached to the large ring, all 36 water inlets at the bottom of the steel tubes were opened by removing the slid valves.

At 11 UTC the large ring was detached from the R/V Pelagia and freely floating. The top-valves were opened starting from the side of drag-parachute (Fig. 1, 10). Five minutes after opening the first valve the entire ring was sinking, the last two tubes being briefly lifted out of the water by several meters. The drag-parachute moved to the center of the large ring and was pulled underwater normally.

A video of the final stage of deployment is available at https://www.youtube.com/watch?v=acFWP1UIWW4

All 6 acoustic releases at the bottom of the drag-parachute lines responded, which implied that the large ring had landed correctly, not upside down. All releases disconnected normally. However, the drag-parachute did not come up to the sea surface, because the two northern lines somehow got caught.

*c. Underwater operation*

For inspection of the unfolded 3D-T mooring array and detachment of the stuck drag-parachute 6000-m rated Remotely Operated Vehicle 'ROV' Victor was launched from Ifremer French R/V Pourquoi pas ? about 6 weeks after deployment, on 19 November 2020. Due to its umbilical cord, the ROV could not be operated inside the large ring and all operations were done via approach along the seafloor from outside the ring.

**5. Results**

The sinking of the large mooring array at 5 min after opening the first top valves was about two times faster than anticipated from the trials with a test-tube in the winter before. During deployment, higher pressure in the steel tubes was noted when opening the valves, with air being forcibly blown out. Presumably, the warmer conditions during deployment compared to the wintertime test make a difference, but it is unknown whether this is the only cause. The faster filling did not affect the anticipated capsizing and sinking towards the side of the drag-parachute.

The twofold triangulation of the six acoustic releases of the drag-parachute after landing of the large ring at the seafloor gave a consistent result to within 5 m precision (Fig. 11). The absolute position is accurate to within about 15 m, which is the approximate horizontal distance between R/V Pelagia's GPS-antenna and the acoustic transducer calling the releases. The 3D-T mooring position at the seafloor is about 80 m East-south-east from its surface position when going underwater.

Two out of six temperature-tilt sensors were lost due to insufficient taping, from releases attached to tubes #9 and #12, which are the closest tubes to the drag-parachute when at the surface and the first ones to go underwater. The remaining four sensors were successfully recovered from the drag-parachute acoustic releases. The temperature-tilt data confirm the initial sideways sinking during the passage of the thermocline about 100 s after going down under, with the sensor attached to tube #7 passing first and the one to #18 last (Fig. 12). During this initial sinking stage tilt varies considerably due to the lost control as in any device or ship sinking, until the tilt rather suddenly becomes constant with time, apart from turbulence vibration noise. This moment of tilt being constant in time implies the line to which the sensor is attached becomes fully stretched. Only when all lines are stretched, the records become steady, which is 300 s after the ring completely went under (and 360 s after #7 went under). This corresponds with the large ring being at approximately 380 m below the sea surface, and the drag-parachute regaining full control over the large mooring at a depth close to the anticipated distance of sinking (Appendix). Only the tilt of #4 continues to vary slightly for another 150 s, possibly due to fact that its line was shortened by a loop around the glass-spheres-acoustic-release packet, as established from the ROV-dive. From then on, about 600 m below the sea surface, the large ring sinks fully controlled horizontally all the way to the 2458 m deep seafloor.

The mean sinking speed of the 3D-T is $1.28\pm0.03$ m s$^{-1}$, as calculated for #18. This value is about 0.1 m s$^{-1}$ lower than expected from model-calculations. The discrepancy is almost entirely attributable to the 15% additional resistance imposed by the drag-parachute. The touchdown at the seafloor is simultaneously to within $\pm1$ s at the four sensors, as far as can be established. This uncertainty is due to vibrations in the drag-parachute lines that take-up a different tilt-attitude after landing, due to the sensors being attached to the freely rotatable and recoverable releases instead of the fixed tubes of the large ring, and due to the sensors not being electrically synchronized. Synchronization was done manually, with difficulty, during post-processing using times of passage through small temperature steps when the large ring was stable.

The tilt of #18 showed a sudden change almost exactly 10 days after deployment, which is expected to have been well after all vertical mooring lines were chemically released (5 to 7 days after being underwater). The sudden change is interpreted as the 'spontaneously' loosening of this stuck release, but the reasons are unknown why after 10 days and how this release was stuck initially. During its loosening it may have (further) damaged some or more sensors from two and possibly four vertical lines in its direction towards #4. The ROV-cutting of the stuck line at #4 tore off and/or moved about 15 T-sensors, all in the upper one-third (above the drag-parachute-position) of the vertical line only 2.5 m away horizontally and to which the drag-parachute leaned.

The ROV-inspection demonstrated that the acoustically opened hook of release at #4 was stuck against its doubling chain. This was caused by the drag-parachute line being looped around the packet of two glass spheres holding the release, whereby the release was tilted with its hook pressed against the doubling chain. It is unknown how this looping occurred and how it could have been prevented. Although the packet could rotate to some extent, the release hook may have been better rotated 180° by unbolting its housing, but that is hindsight information and speculation.

The ROV has monitored about half the number of the vertical lines from outside the large ring. All visible were upright and well-positioned as in (Fig. 13) indicating that the chemical releases had worked properly. The 12 mm cable network seems well-stretched and is suspended from the seafloor. The only damage, apart from the lost 15 T-sensors, seems the deterioration of the zinc-aluminum-alloy anodes mounted on the large ring (Fig. 14). During the large-ring-assembly in the harbor fast oxidation was already noted in the anodes. ROV touching one anode showed basically complete powdering. Hence, the electro-chemical protection of the steel tubes is expected to be soon lost, much faster than anticipated by the manufacturer of the anodes. Previous similar experiences with anchors of ANTARES lines, now 10 years underwater, showed no damage to the steel (J. Brunner, pers. comm. 20 November 2020). It is anticipated that the large amount of steel will ensure the 3D-T steel tubes survive the 2.5 to 3 years of intended mooring underwater.

As expected, the eight corner small rings are tilted due to pull of the three support steel cables on the main steel cable network (Fig. 15). While the image gives a false impression of the instrumented cable being not vertical, which is not the case, the estimated tilt of the small ring is between 15 and 20° from the horizontal. As a result, the vertical line is displaced higher by about 0.35 m than would be if mounted in the steel cable network at 0.3 m above the seafloor at this position. As far as can be judged, the steel cable network is tightened and tilted from the seafloor as theoretically computed.

The area where the 3D-T mooring landed is indeed very flat, with fluffy material and yet a solid seafloor with some but not excessive imprint between 0.05 and 0.1 m in the sediment of the tubes of the large ring (e.g., Figs 14, 15). Larger than expected are the amount of debris on the seafloor and the number of bugs, small animals that are particularly active as soon as the seafloor material is resuspended (Fig. 16).

## 6. Conclusions

The assembly and deployment operation of the 3D-T mooring array was very satisfactorily mechanically, even though the drag-parachute got stuck possibly due to a combination of a line wrapping around the glass spheres and orientation of the hook towards its doubling chain.

The planned sinking and regaining of control using a parachute adding 15% drag to the large ring worked well. The filling of the steel tubes by opening mechanical valves at bottom first and top last proved control over the side of sinking and allowed ample time to fill the tubes with water and prevent implosion during the sinking to great depths. This completes the advantage of using the large ring as a floating device, for easy towing, and as an anchoring device, for vertical mooring lines at the seafloor. This 'controlled' free-fall deployment resulted in a smooth and horizontal landing of the large ring at the seafloor at a vertical speed of 1.3 m s$^{-1}$. All visible vertical mooring lines are well deployed via their chemical releases. We are looking forward to recover these lines via deep-sea submersible and the high-resolution data from the nearly 3000 T-sensors in 2.5 to 3 years after deployment.

For future deployments of different, complex mooring structures, we thus recommend the use of a sturdy parachute with drag at sinking speed constructed about half by buoyancy elements and half by surface area.

We also recommend to verify the required sinking time under realistic conditions and moderate valve openings if necessary. We did our tests under cold winter conditions, not the warm, sunny autumn conditions that built higher pressure in the tubes when deploying.

*Acknowledgments.* This research was supported in part by NWO, the Netherlands Organization for the advancement of science. We thank captain Len Bliemer and the crew of R/V Pelagia and captain Gilles Ferrand and the crew of Pourquoi pas ? for the very pleasant cooperation. We also thank Franck Rosazza and his team of ROV Victor for the well-performed underwater mission. NIOZ colleagues notably from NMF-department are thanked for their indispensable contributions during the long preparatory and construction phases to make this unique sea-operation successful. Jesper van Bennekom made the schematic drawings of Fig. 5, Rob Buiter took the photo of Fig. 6. Underwater images of Figs 13-16 are stills of video recordings from ROV Victor (Ifremer, France).

**Appendix. Calculation of friction during fall for actual drag-parachute**

The principle of sinking is a balance between gravity and drag-resistance.

We have resistance (pulling) force: $F_w = ½ρC_DAW^2$, underwater 'weight' $F_g = G - F_A$ that corrects weight $G = mg$ for Archimede force $F_A$, and braking force $F_b = F_w + B$, B the buoyancy or floating force, where m is the mass, g is the gravitational acceleration, ρ is the density of water, $C_D$ and A the drag coefficient and the surface in the direction of fall of the object, respectively, W the fall-speed. For the density of seawater we use $ρ = 1026$ kg m$^{-3}$ and we take $C_D = 1$ for a cylinder.

The total large ring mooring array has $F_g = 177$ kN. The large ring has an area of 0.61x11.9x18 = 130.7 m$^2$. Assuming that a small ring including top-buoy has 1 m$^2$ area, the array has a total area of about 175 m$^2$. So the estimated fall speed during balance but without drag-parachute becomes:

$W = (177000×2/1026/1/175)^{1/2} = 1.4$ m s$^{-1}$.

For the scale model, calculations have been made using a speed of 1.5 m s$^{-1}$, which is close to the above value.

Taking 15% of $F_g$, the actual drag-parachute must deliver $F_b = 0.15F_g = 26.2$ kN of braking force. By using floatation with a total buoyancy of $B = 15.9$ kN, the disk must deliver $F_w = 10.3$ kN of resistance force at 1.4 m s$^{-1}$. That implies an area of:

$A = 10300×2/1026/1/1.96 = 10.2$ m$^2$,

or a disk with a diameter of 3.6 m.

In the final construction, this resistance is distributed over the large brake disk (weighing 2.1 kN under water) with surface area of $A = 9.6 +$ ring $= 10$ m$^2$ (and 48 glass spheres providing $B = 12.2$ kN) and $A = 1.7$ m$^2$ from the additional standard mooring top-buoy (providing $B = 3.7$ kN). This totals to $F_b = 25.6$ kN $= 0.145F_g$.

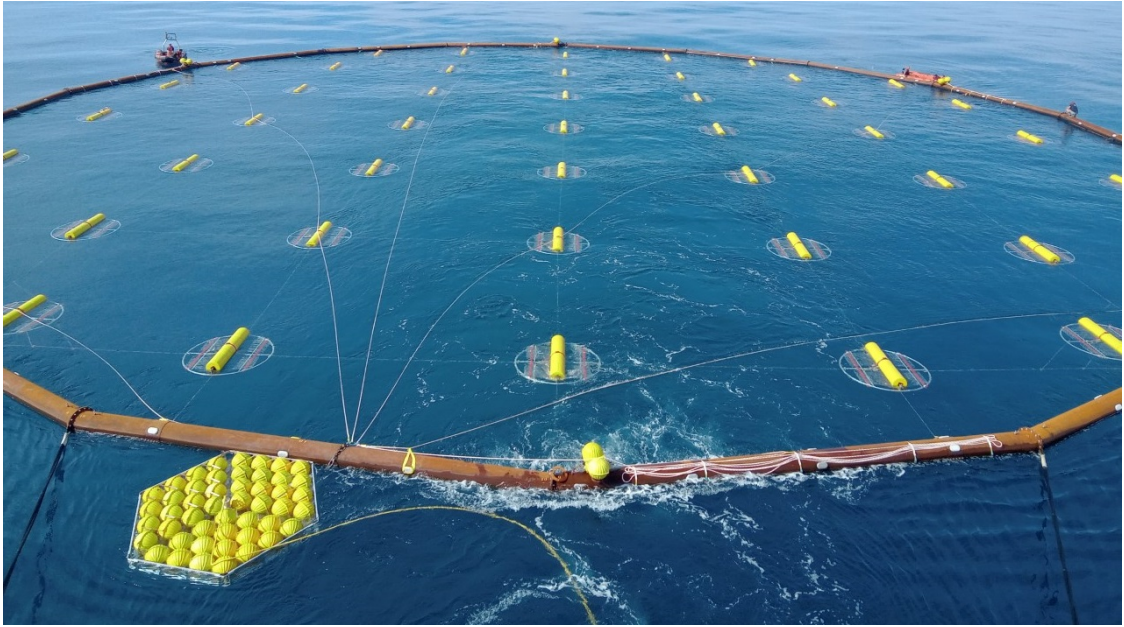

FIG. 1. The entire large-ring mooring array '3D-T' in compacted form just before deployment. Visible are most of the 45 buoys on top of the small rings in which the mooring lines are fold-up with T-sensors. The small rings are attached to a steel cable network that is tensioned between the steel tubes and which is vaguely visible underwater. Outside the large ring, the drag-parachute is attached in the front-left. The drag-parachute is connected to 6 white floating lines of which the nearest is attached in loops on the large-ring steel tube in the front. The two yellow balls contain glass spheres to underwater lift the acoustic release between them and to which the white line is attached, for removal of the drag parachute immediately after landing on the seafloor.

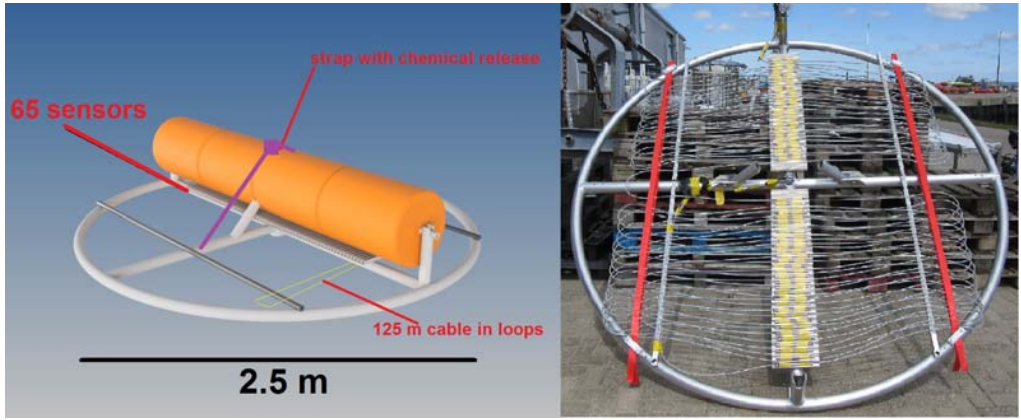

FIG. 2. The small ring assembly. (a) Schematic with cable channels for 65 T-sensors and buoy on top. (b) Small ring ready for transportation with 125 m mooring line holding T-sensors looped and held in cable channels, straps and clips.

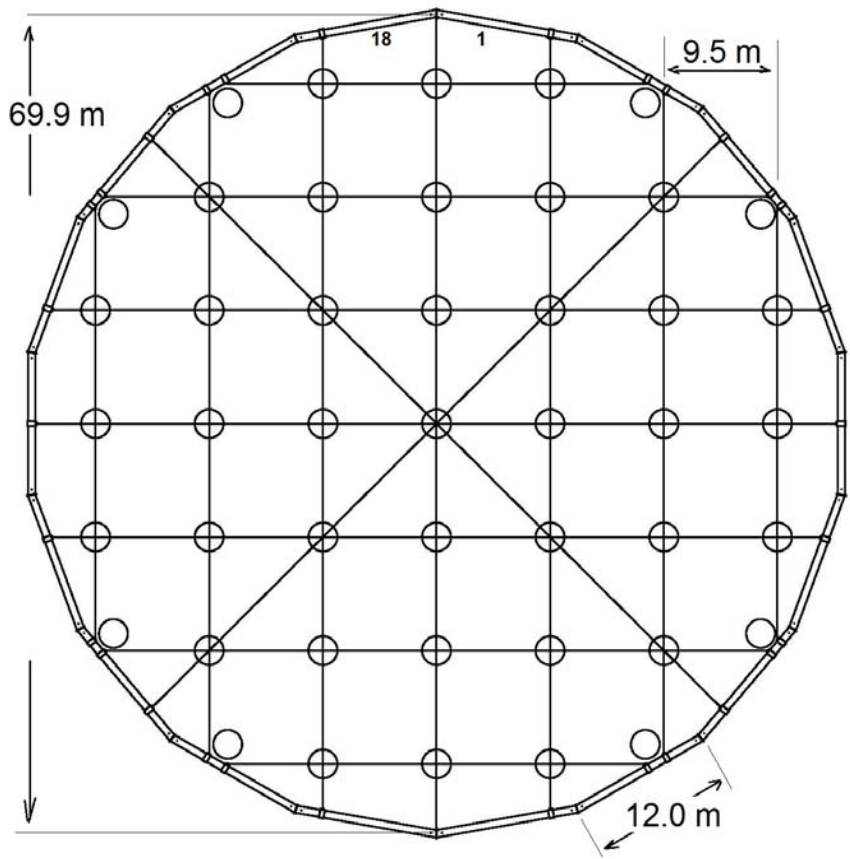

FIG. 3. Floorplan dimensions of the large ring with its 12 mm steel cable network and small rings at intersections of the network. Steel tube numbering is clockwise from 1 to 18 as viewed from above.

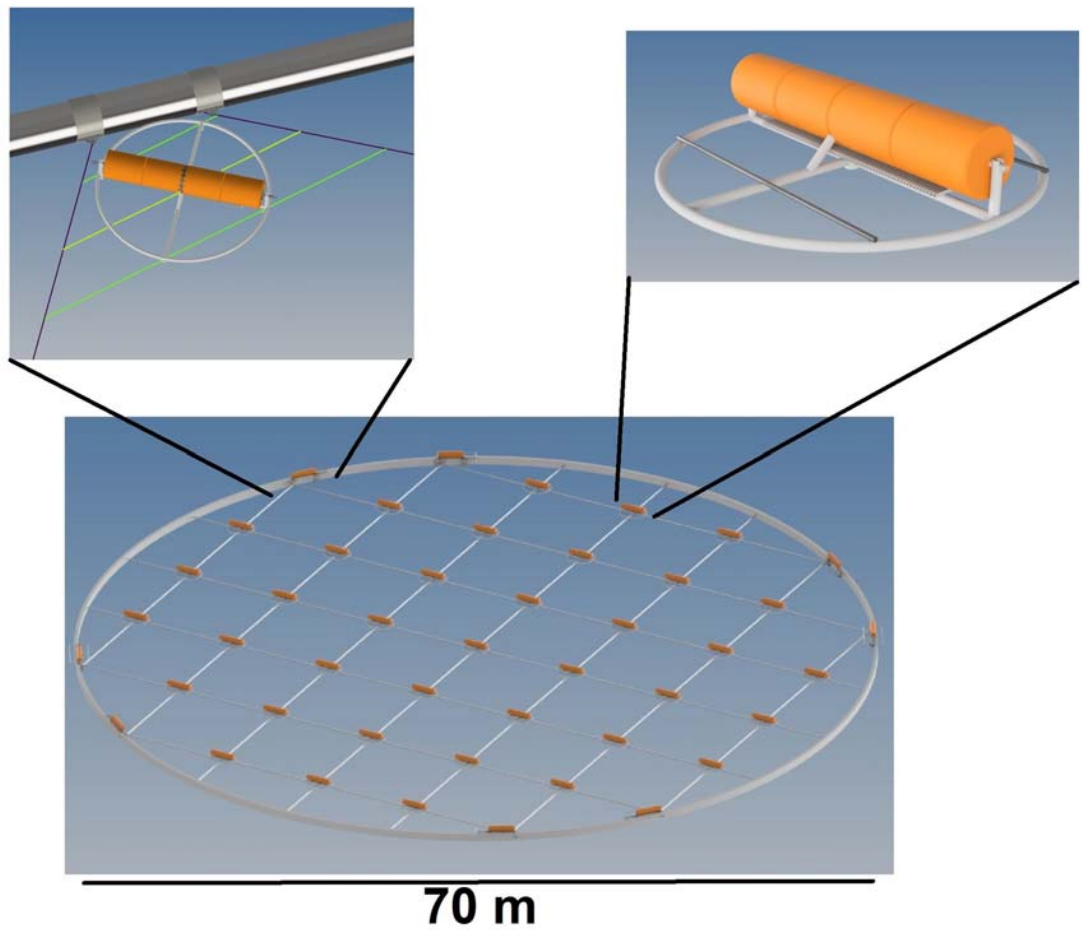

FIG. 4. Schematic of the large ring with small ring mounting in a corner (upper left) and at an intersection (upper right) of the 12 mm steel cable network.

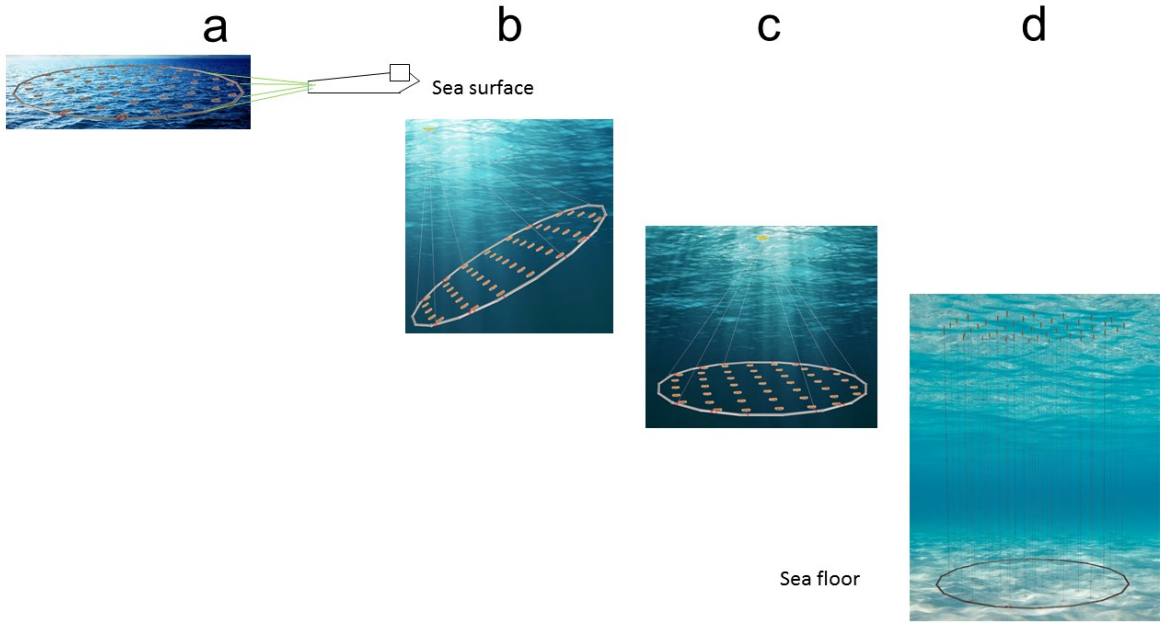

FIG. 5. Schematic for the sinking of the large 3D-T mooring array. The lighting in the sea under water are artistic freedom as it is pitch-dark below about 700 m. (a) Towing-out in flat state from harbor to mooring site. On-site attachment of the drag-parachute and opening of valves to fill the large ring with seawater. (b) The initial uncontrolled stage of underwater sinking with the drag-parachute near the surface. (c) Deeper down, the drag-parachute has gained back control and balances the large ring into a steady state. (d) After 5 to 7 days at the seafloor, mooring lines appear vertically after resolution of chemical releases liberating the top-buoys.

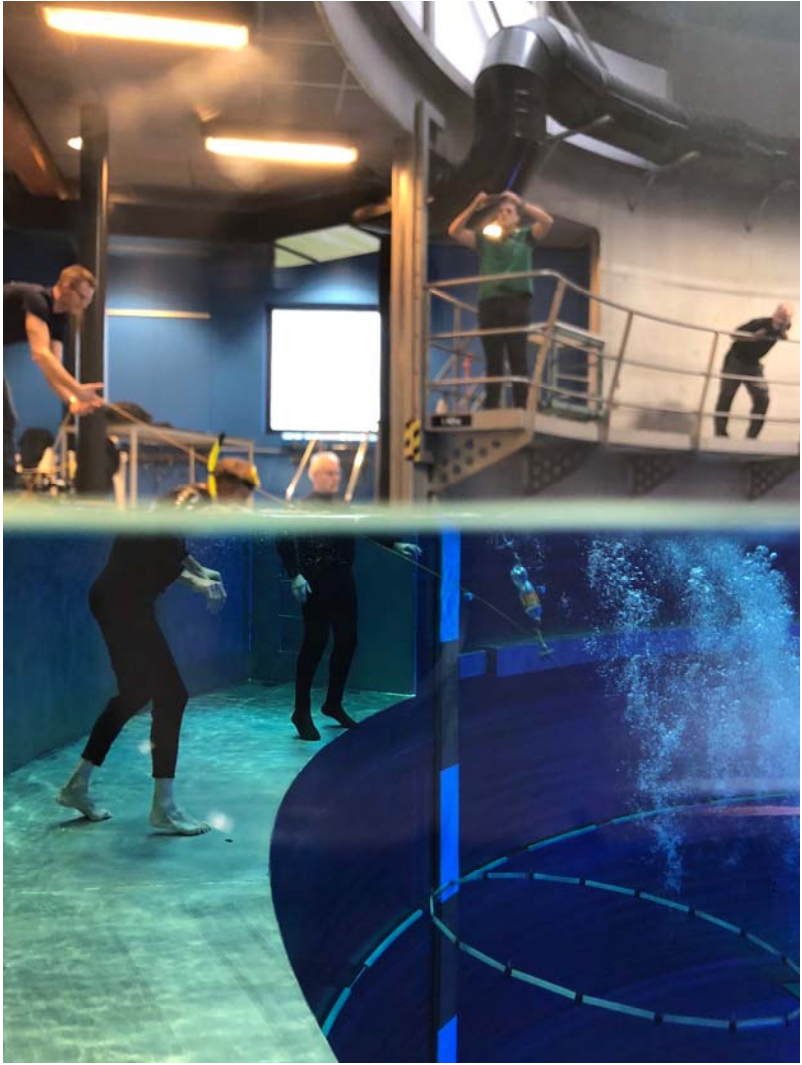

FIG. 6. Impression of testing the drag-parachute using scale models in an 8 m deep swimming pool. The large ring steel model is in the lower right corner, the drag-parachute model just below the water surface to the left of the air-bubble columns.

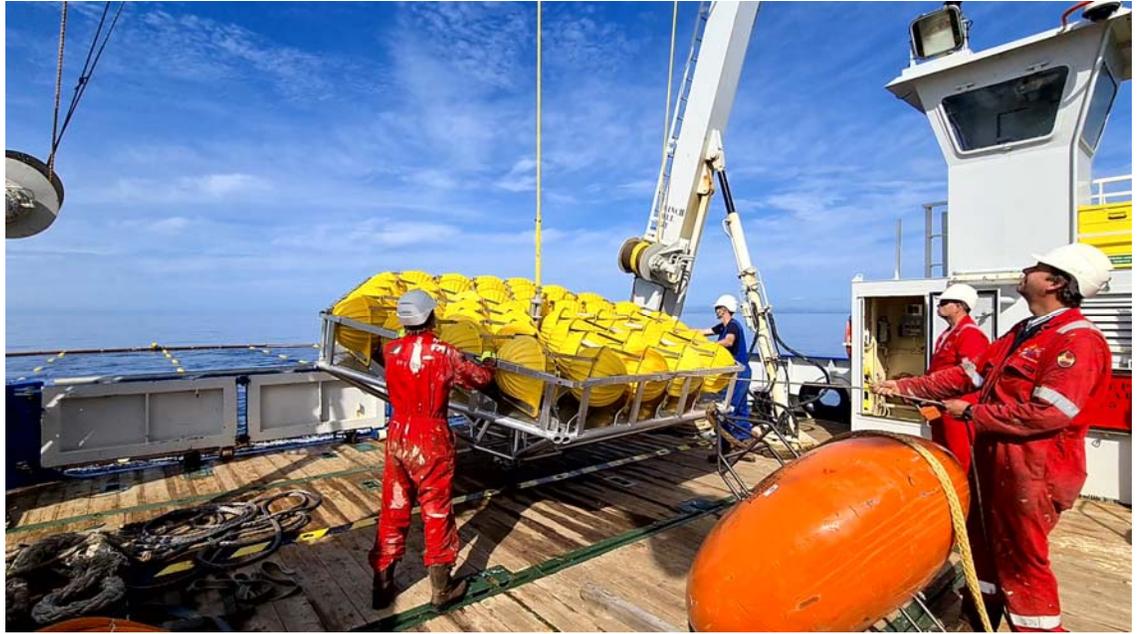

FIG. 7. Custom-made drag-parachute being hoisted, with orange standard elliptic buoy on deck to the right.

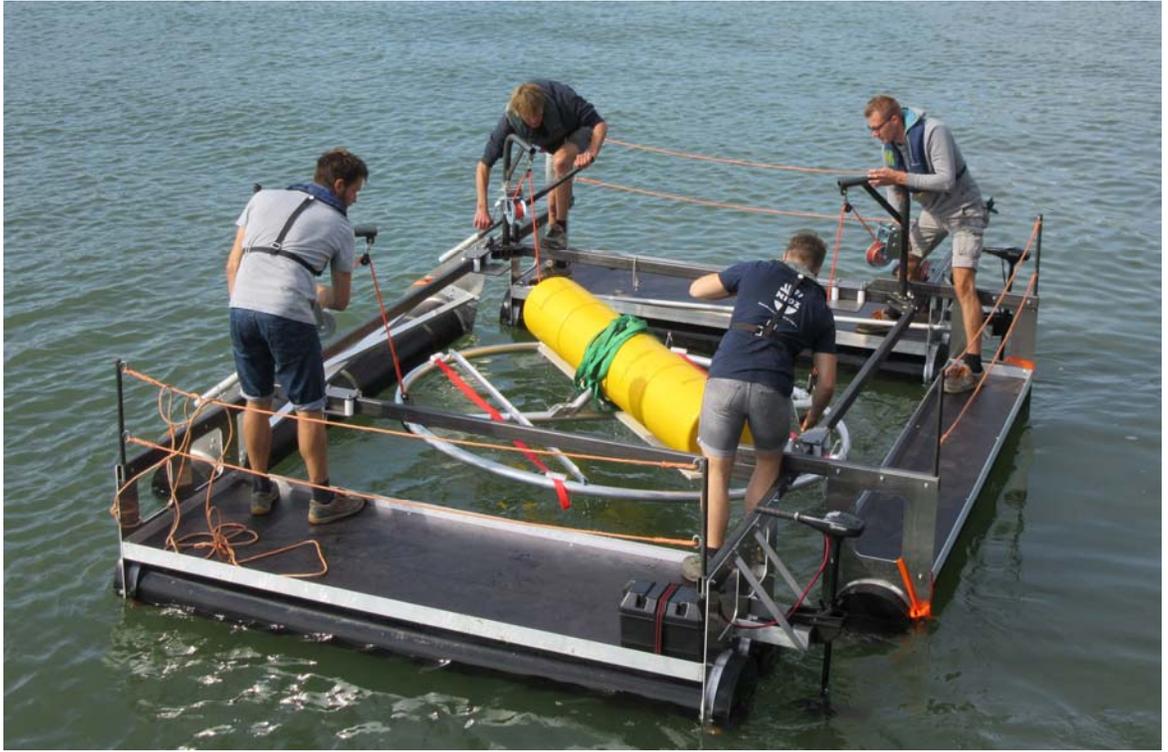

FIG. 8. Test of the small-rings-to-cable-network assembly raft in the NIOZ-harbor.

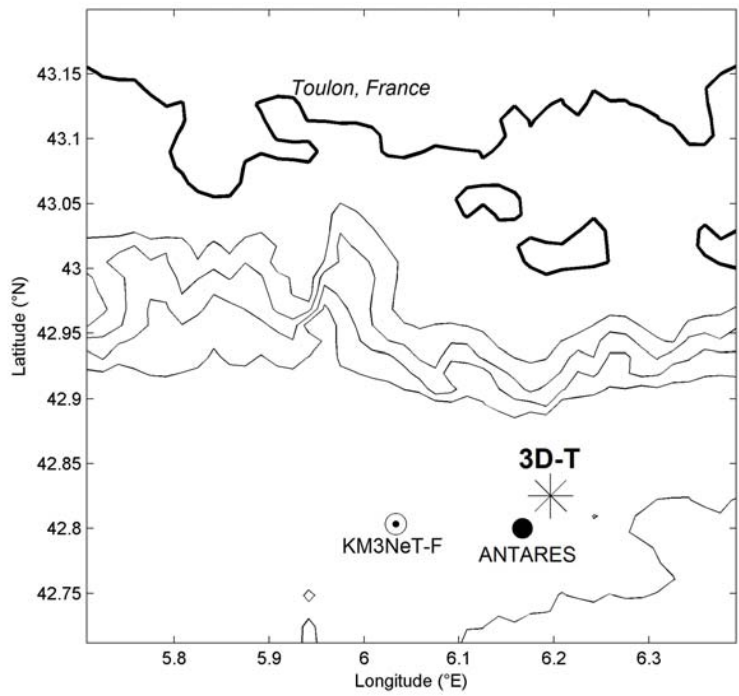

FIG. 9. Location of the 3D-T mooring (star), just inside the French 12-miles zone of territorial waters. The two neutrino telescope sites are in the vicinity. Bathymetry is from the Smith and Sandwell (1997) database, the 1-km version 9.1b, with depth contours given every 500 m.

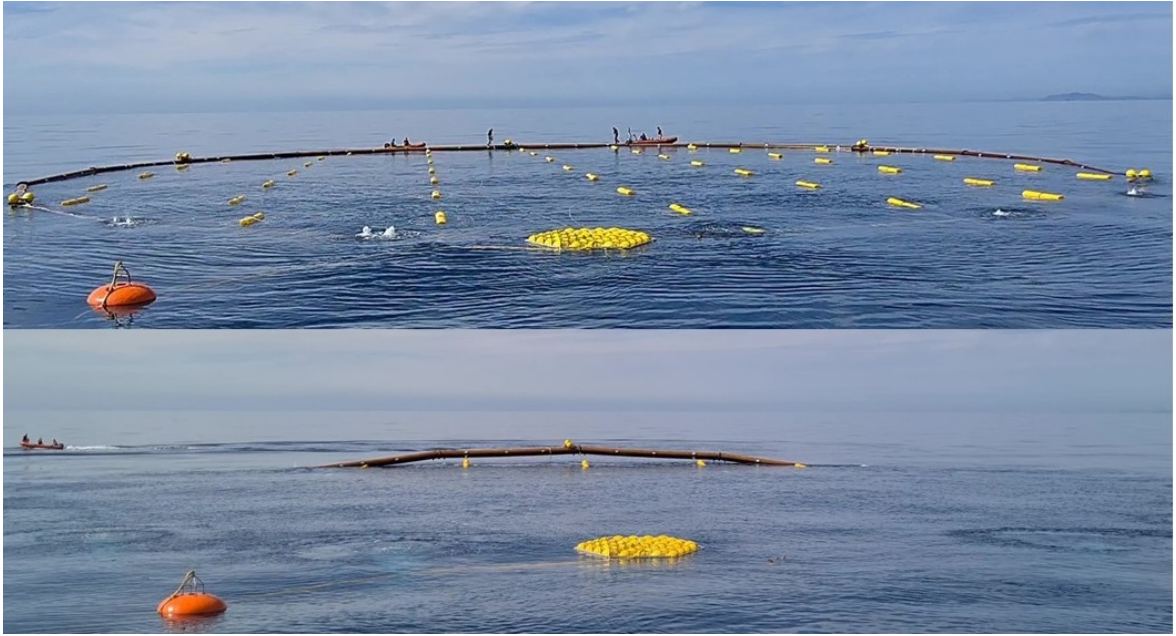

FIG. 10. Two video stills from the final minute of valve opening before sinking of the

  3D-T mooring array.

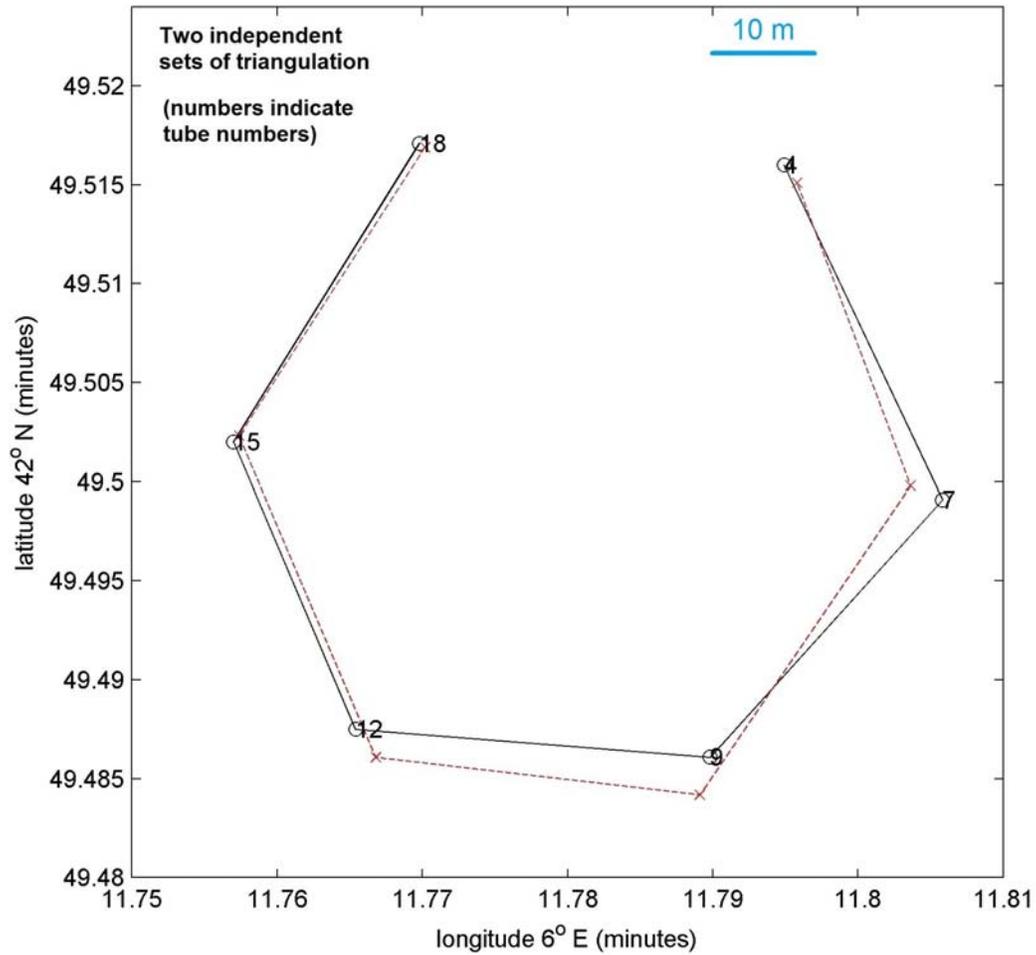

FIG. 11. Triangulation using acoustic releases at the bottom of 6 drag-parachute lines for positioning of the large ring at the seafloor. Two independent sets of data are used, but no correction is made for the difference in horizontal position of the acoustic transponder with respect to R/V Pelagia's GPS antenna. The local water-depth is 2458 m. The numbers refer to the tubes to which the acoustic releases were attached.

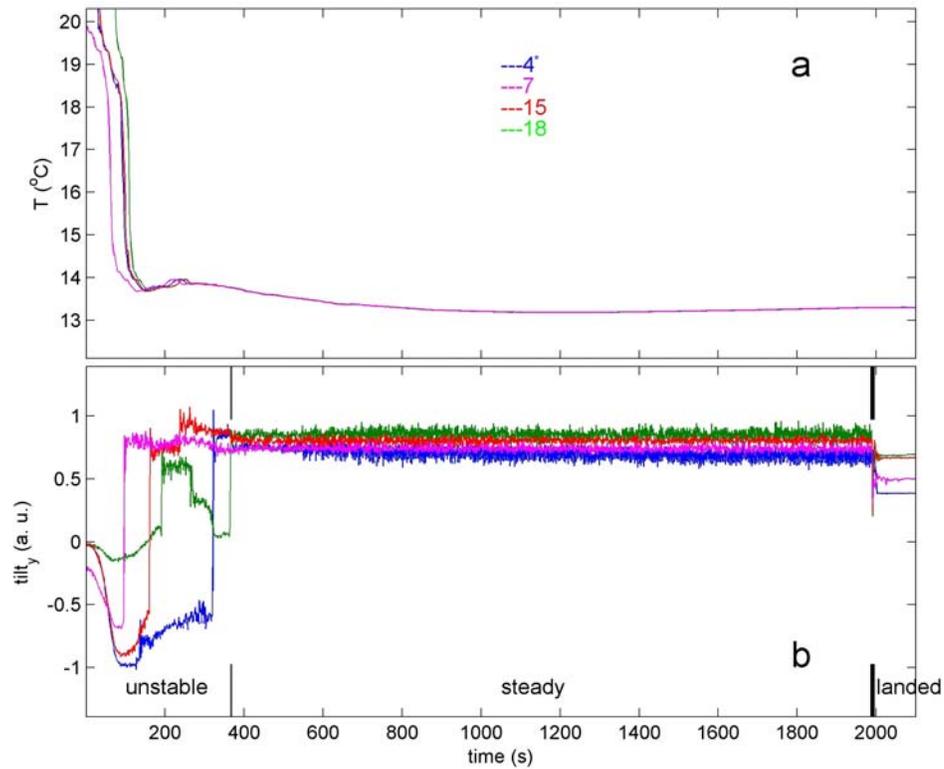

FIG. 12. Sinking information as a function of time after tube #7 going underwater until landing at the seafloor from four temperature-tilt sensors mounted to acoustic releases on drag-parachute lines attached to large-ring tubes (cf. Fig. 11 for their orientation). (a) Temperature. The numbers refer to the tubes to which the lines were attached and of which #18 (spontaneously loosened after 10 days) and #4 were stuck, of which the line of #4 was shortened due to looping around the glass-sphere-acoustic-release packet. (b) Normalized tilt for the y-component. The left thin vertical lines indicate when the ring becomes steady and tilt constant with time. The right solid vertical lines indicate touchdown at the seafloor.

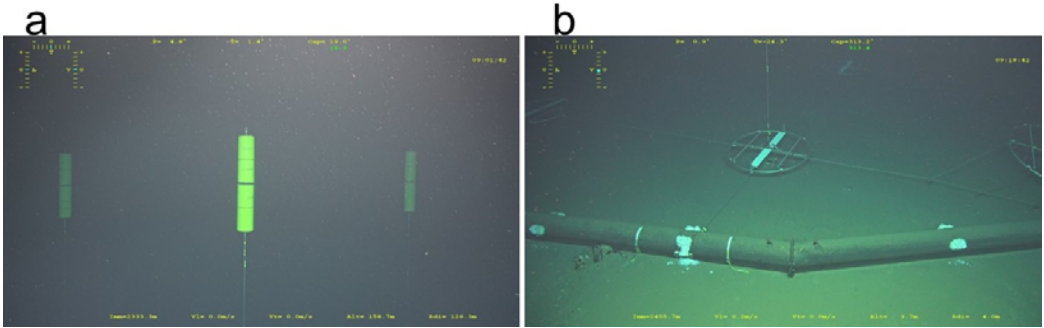

FIG. 13. Video-stills of underwater ROV-inspection of well deployed vertical mooring lines with temperature sensors. (a) Top-buoys at about 125 m from the seafloor. (b) Two tubes of the large ring in the foreground and several small rings with tight steel cable network.

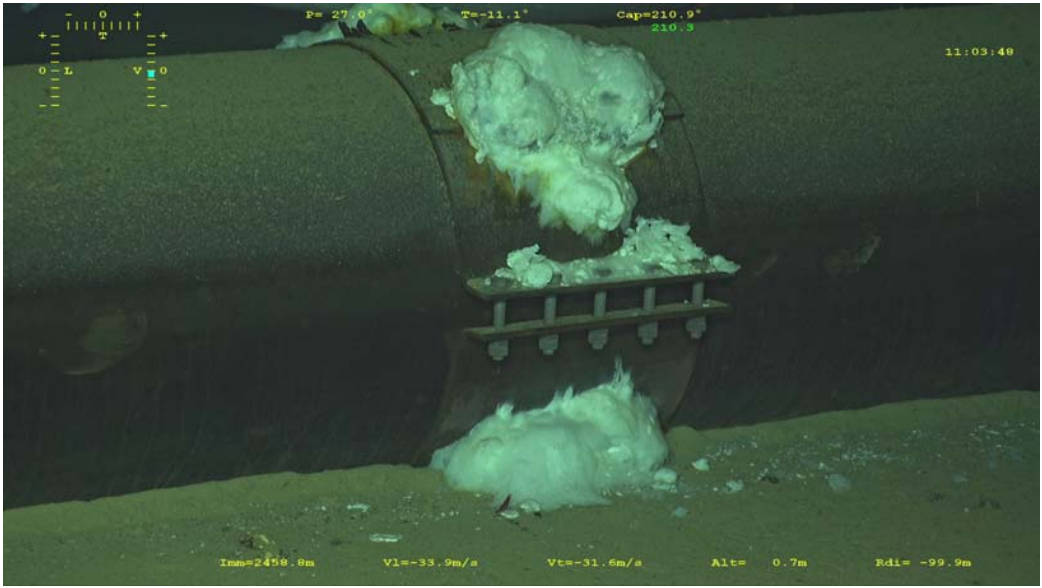

FIG. 14. Detail of cable attachment saddle with heavily deteriorating zinc-aluminum-alloy anode.

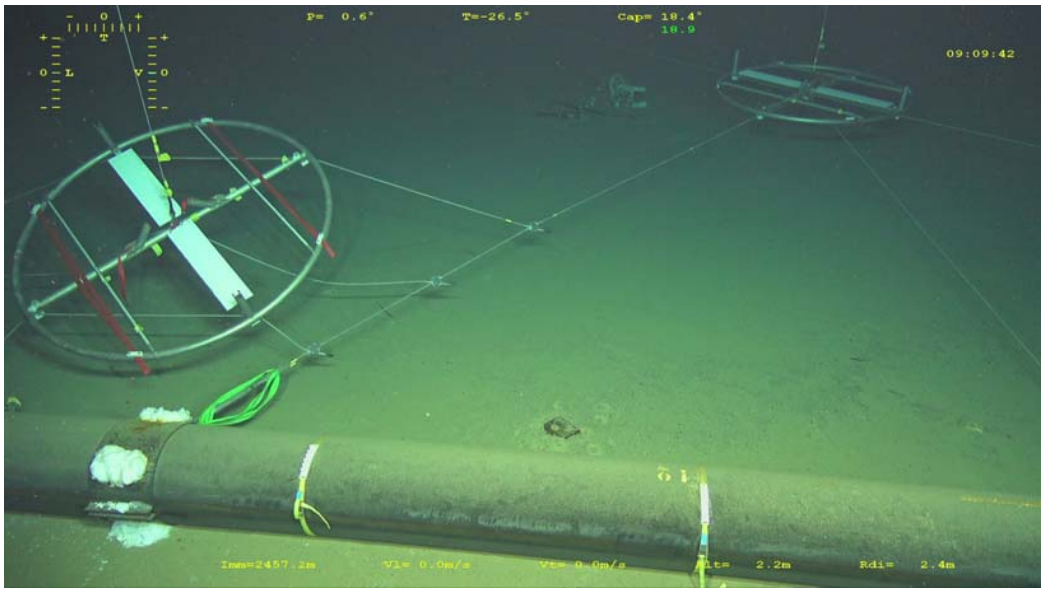

FIG. 15. Detail of a tilted corner small ring after unfolding of its vertical line.

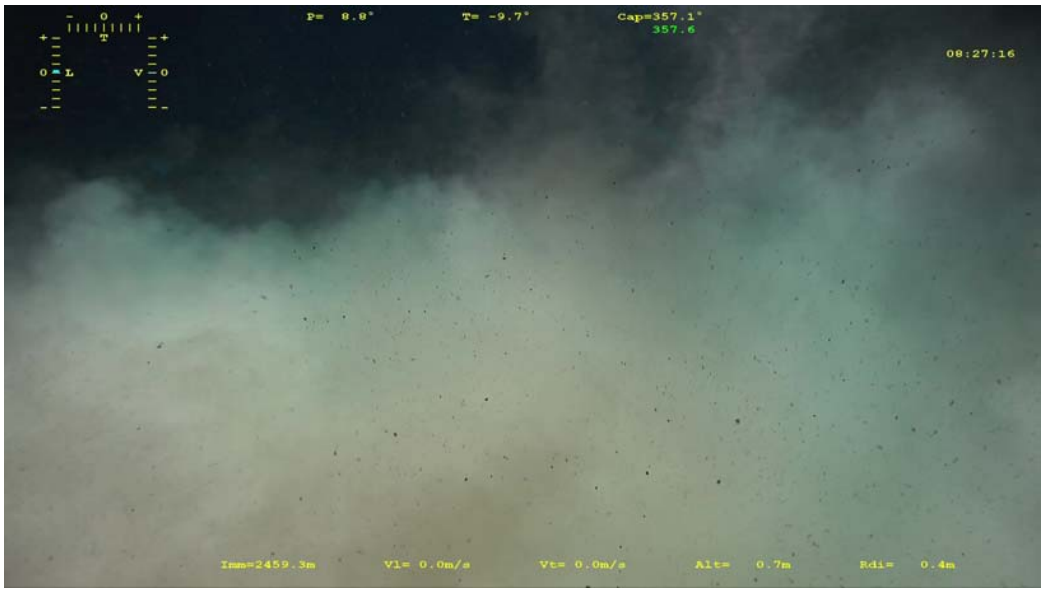

FIG. 16. ROV-inflicted dust cloud with numerous small animals (black spots) after unintended seafloor touchdown.